\documentclass[aps,pra,superscriptaddress,twocolumn,showpacs,noeprint]{revtex4-2}
\usepackage{graphicx,amsfonts,times,bm,amsmath,verbatim,color,array}

\usepackage{graphicx,amssymb,amsmath,ifthen,braket,xcolor}
\usepackage{bm}

\begin{document}

\title{Stabilizing Topological Superfluidity of Lattice Fermions}

%==========================================================================================
\author{Junhua Zhang}
\affiliation{Department of Physics, Virginia Tech, Blacksburg, Virginia 24061, USA}
\author{Sumanta Tewari}
\affiliation{Department of Physics and Astronomy, Clemson University, Clemson, South Carolina 29634, USA}
\author{V.W. Scarola}
\affiliation{Department of Physics, Virginia Tech, Blacksburg, Virginia 24061, USA}

\begin{abstract}
Attractive interaction between spinless fermions in a two-dimensional lattice drives the formation of a topological superfluid. But the topological phase is dynamically unstable towards phase separation when the system has a high density of states and large interaction strength. This limits the critical temperature to an experimentally challenging regime where, for example, even ultracold atoms and molecules in optical lattices would struggle to realize the topological superfluid. We propose that the introduction of a weaker longer-range repulsion, in addition to the short-range attraction between lattice fermions, will suppress the phase separation instability.  Taking the honeycomb lattice as an example, we show that our proposal significantly enlarges the stable portion of the topological superfluid phase and increases the critical temperature by an order of magnitude.  Our work opens a route to enhance the stability of topological superfluids by engineering inter-particle interactions. 
\end{abstract}

\maketitle

\section{Introduction}

Interest in engineering topological superfluids stems from fascinating aspects of their excitations \cite{Moore1991a,Read2000,Nayak2008}.  When spinless fermions pair, the resulting superfluid can have a chiral $p$-wave ($p_x+ip_y$) Bardeen-Cooper-Schrieffer (BCS) order parameter.  Vortices in such two-dimensional (2D)  superfluids are predicted to host Majorana fermion zero modes and therefore display non-Abelian braid statistics \cite{Ivanov2001}.  If observed, these topological superfluids and related vortices could have potential applications in  topological quantum computing \cite{Nayak2008}.

Atoms and molecules placed in optical lattices allow quantum state engineering with accurately tunable parameters and, as a result, have had tremendous success in realizing quantum many-body phases of matter \cite{Lewenstein2007,Bloch2008}.  They are therefore prime candidates for realizing topological superfluids. Several proposals to introduce attractive interactions between ultracold fermionic atoms or molecules hope  to create 2D topological superfluids via: $p$-wave Feshbach resonances \cite{Gurarie2007}, dipolar moments \cite{Baranov2012,Gadsbolle2012,Lin2013}, synthetic spin-orbit coupling \cite{Zhang2008a,Sato2009}, dissipation \cite{Bardyn2012}, orbital effects \cite{Buhler2014}, and atomic gas mixtures \cite{Mathey2006,Massignan2010,Wu2016,Midtgaard2016}.  However, it remains challenging to observe the phase experimentally, partly due to its rather low critical temperature.

Increasing the stability of a topological superfluid in a lattice is not as straightforward as it first appears.  Increasing the attractive interaction strength $V$ and increasing the density of states at the Fermi level $\rho$ are well known routes  to raising the critical temperature $T_c$ of a typical BCS state since $T^{\text{BCS}}_c\sim\exp{(-1/V\rho)}$.  Related theoretical studies \cite{Cheng2010} of spinless fermions on 2D lattices at first confirmed this by showing that a highly robust topological superfluid phase is  energetically favorable in a large family of lattice models therefore offering hope for realizing these states in the laboratory. However, further studies showed that the system becomes dynamically unstable towards phase separation, indicated by a negative inverse compressibility, at high densities and large interactions \cite{Liu2012,Midtgaard2016}.  The phase separation instability that occurs at a high density of states and with large attractive interactions between spinless lattice fermions hinders prospects for achieving higher temperature topological superfluidity. 

We use Hartree-Fock theory to systematically study spinless fermions with just nearest-neighbor (NN) attraction on various 2D lattices.  We find that the unwanted competition between a uniform pairing phase and phase separation is common.  The combination of particle-hole duality and a renormalization of the chemical potential through the Hartree contribution of the attractive interaction causes phase separation. This Hartree effect is generic to nearly any lattice and is more pronounced at high density of states, for instance, near a Van Hove singularity, and at large attraction.  But a high density of states and a stronger attraction are, as argued above, precisely what is needed to obtain a higher superfluid critical temperature. We therefore see that the mechanisms needed to stabilize topological superfluidity induce a strong Hartree effect such that the high-density superfluid phase is dynamically unstable and separates into a mixture of two low-density (one low particle-density and one low hole-density) superfluids, associated with lower critical temperature.  Phase separation therefore poses a significant obstacle to realizing topological superfluids with atoms or molecules in optical lattices.

We propose that while short-range attraction in a lattice can induce a superfluid, a weaker longer-range repulsion enhances the critical temperature of the superfluid by suppressing phase separation in a mechanism similar to the one studied in the context of the $t-J$ model \cite{Pryadko2004}.  Interactions with such an alternating sign (attractive at short-range but repulsive at long-range) akin to an Ruderman–Kittel–Kasuya–Yosida (RKKY) interaction \cite{VanVleck1962} have been examined in the context of ultracold atoms and molecules.  For example, an alternating sign interaction was discussed in the context of multispecies boson mixtures \cite{Soyler2009}.  Experiments have also realized a related RKKY-type interaction with Bose-Fermi mixtures \cite{Edri2020}.   Another promising route to engineering an interaction with an alternating sign has been discussed in the context of microwave dressed states of polar molecules \cite{Baranov2012}.  

We use unrestricted Hartree-Fock theory to examine a minimal model of interacting spinless fermions on a lattice.  Recent work shows that phase diagrams produced using this approximation compare well with exact diagonalization results when applied to models of this type \cite{Hui2018,Chen2018}.  We study the honeycomb lattice as a demonstration.  Our results are consistent with exact diagonalization results on the honeycomb lattice at half filling \cite{Capponi2015,Capponi2017} where we do not expect a superfluid.  We use our method to go on to study lower densities where attraction between NNs on the lattice leads to a topological superfluid.  The critical temperature is found to be very low, three orders of magnitude below the lattice bandwidth, as expected from the unwanted Hartree effect.  We then include longer-range repulsion to find an order of magnitude increase in the critical temperature.  We therefore explicitly demonstrate the following mechanism: the addition of a weaker longer-range repulsion to the otherwise attractive inter-particle interaction enhances topological superfluidity by suppressing phase separation.  Our work sets the stage to examine the use of interactions with alternating signs to stabilize topological superfluids.

The paper is organized as follows.  In Sec.~\ref{sec_model} we discuss an interacting model of spinless fermions on the honeycomb lattice and the unrestricted Hartree-Fock method used to solve it.  Sec.~\ref{sec_phases} overviews the competing orders that arise in our model.  These include the topological superfluid, a normal metal, a zigzag stripe phase, and phase separation.  Sec.~\ref{sec_enhancing} presents the results of our calculation.  Here we show that, without a weaker longer-range repulsion, attractive fermions form a topological superfluid only at low temperatures because of competing phase separation.  We then show that the addition of longer-range repulsion enhances the stability of the topological superfluid.  Here we optimize parameters and compute the phase diagram to reveal where the topological superfluid is most stable.  We conclude in Sec.~\ref{sec_conclusion}. 

\section{Model and Methods}
\label{sec_model}

We consider spinless fermions hopping on a 2D lattice captured by a Hamiltonian:
\begin{align}
H=H_0+H_{\text{int}},
\end{align}
with a non-interacting part:
\begin{align}
    H_0=-t\sum_{\langle ij\rangle}\left(c^\dagger_{i} c^{\vphantom{\dagger}}_{j}+\mathrm{H.c.}\right)-\mu\sum_{i}c^\dagger_{i} c^{\vphantom{\dagger}}_{i},\label{eq_H0}
    \end{align}
    and an interacting part:
    \begin{align}
    H_{\text{int}}=\frac{1}{2}\sum_{i\neq j}V_{i,j}c^\dagger_{i}c^\dagger_{j}c^{\vphantom{\dagger}}_{j}c^{\vphantom{\dagger}}_{i},
    \label{eq_Hint}
\end{align}
where $c^\dagger_{i}$ ($c^{\vphantom{\dagger}}_{i}$) creates (annihilates) a fermion at site $i$, $t>0$ is the hopping amplitude, $\langle ij \rangle$ represent the bonds that connect NN lattice sites $i$ and $j$, $\mu$ is the chemical potential, and $V_{i,j}$ denotes the interaction between fermions at sites $i$ and $j$.  For the interaction, we consider a minimal $V_1$-$V_2$ model with attractive NN interaction $V_1<0$ and repulsive next nearest neighbor (NNN) interaction $V_2\geq 0$:
\begin{align}
    H_{\text{int}}=\frac{V_1}{2}\sum_{\langle ij \rangle}c^\dagger_{i}c^\dagger_{j}c^{\vphantom{\dagger}}_{j}c^{\vphantom{\dagger}}_{i}+\frac{V_2}{2}\sum_{\ll ij\gg}c^\dagger_{i}c^\dagger_{j}c^{\vphantom{\dagger}}_{j}c^{\vphantom{\dagger}}_{i},
\label{eq_H_V1_V2}
\end{align}
where $\langle\langle ij \rangle \rangle$ denotes bonds connecting NNN lattice sites.  In the following we work in units with $k_{\text{B}}=1$. We also set the NN lattice spacing to unity.

%%%%%%%%%%%%%%%%%%%%%%%%%%%%%%%%%%%%%%%%%%%%%%%
\begin{figure}[ht]
	\begin{center}
	   %\vspace{1cm}
		\includegraphics[width=0.48\textwidth]{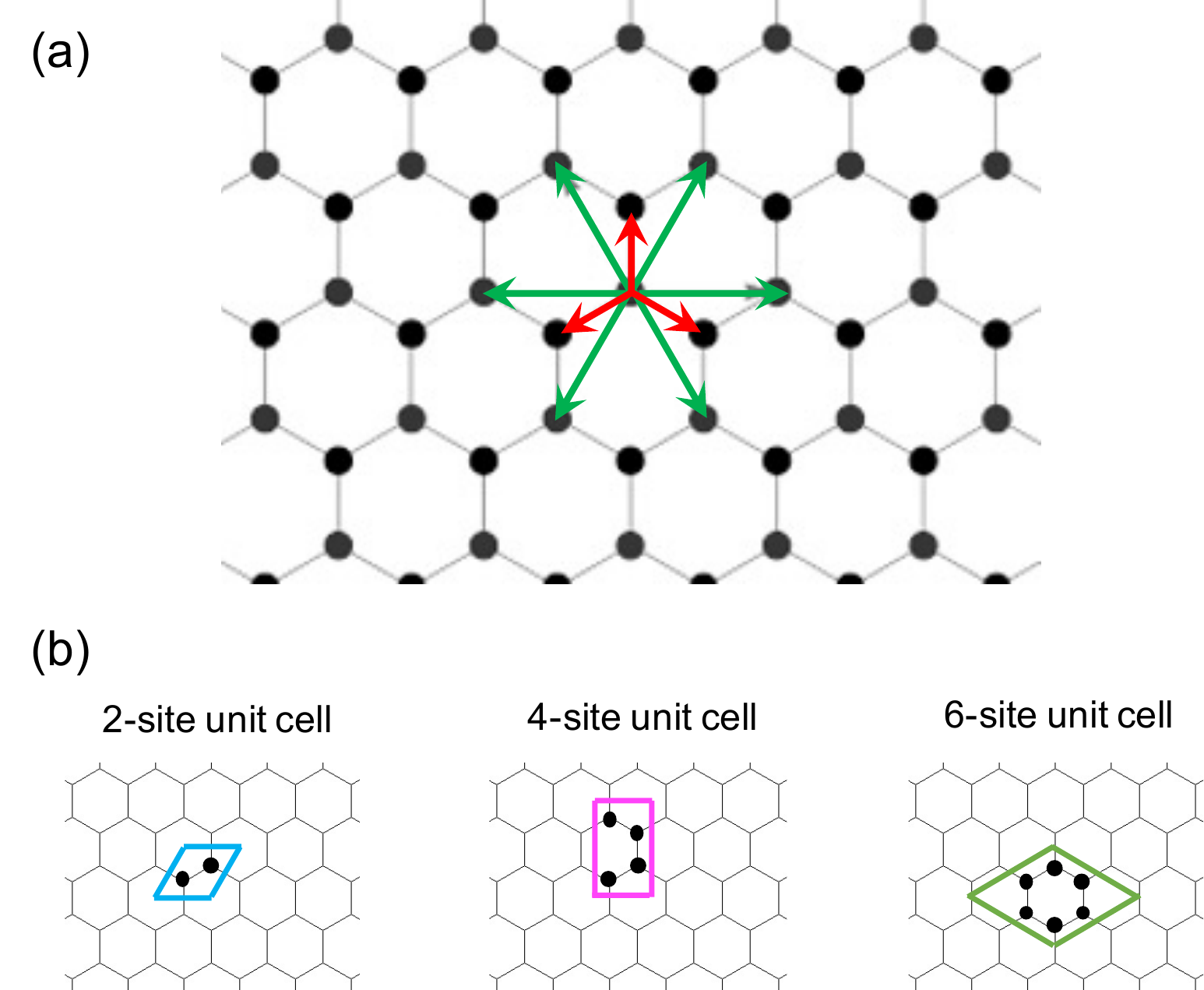}
	\end{center}
	\caption{(a) Schematic of a honeycomb lattice.  The red and green arrows denote the three nearest-neighbor and six next-nearest-neighbor bonds emanating from a central lattice site, respectively.  In our model, spinless fermions have a nearest neighbor attraction, $V_1<0$, and next nearest neighbor repulsion, $V_2\geq 0$.   (b) Choices of the unit cells used in our self-consistent mean-field calculations with two, four, and six sites.}
	\label{fig_schematics_honeycomb}
\end{figure}
%%%%%%%%%%%%%%%%%%%%%%%%%%%%%%%%%%%%%%%%%%%%%%%

We consider fermions hopping on a honeycomb lattice, Fig.~\ref{fig_schematics_honeycomb}.  The red arrows denote NN bonds and the green arrows denote NNN bonds.  The non-interacting band structure of the particles hopping between NNs on the honeycomb lattice, Eq.~\eqref{eq_H0}, has two bands with particle-hole symmetry with respect to the Dirac points at the Brillouin zone corners which lie at the Fermi level when the system is half-filled, i.e., $n=0.5$.  Here $n$ is the fermion number density (or filling fraction) which is the average fermion number per site:
\begin{align}
    n=\frac{1}{N}\sum_i\langle c^\dagger_{i} c^{\vphantom{\dagger}}_{i}\rangle,
\label{eq_n}
\end{align}
where $N$ is the total number of lattice sites.  The density of states vanishes at the Dirac points. On the other hand, the bands have Van Hove singularities at the filling fractions 3/8 and 5/8, and the band character switches between particle and hole band across the Van Hove fillings. Interaction effects can be significantly enhanced near the Van Hove fillings due to the large density of states.

We treat the interaction term, Eq.~\eqref{eq_Hint}, using the unrestricted Hartree-Fock approximation.  It can be decoupled into three channels: the Hartree, Fock, and pairing channels. Introducing the mean fields: $\bar{n}_{i}=\langle c^\dagger_{i} c^{\vphantom{\dagger}}_{i}\rangle$, $\psi_{ij}=\langle c^\dagger_{i} c^{\vphantom{\dagger}}_{j}\rangle$, and $\Delta_{ij}=\langle c^{\vphantom{\dagger}}_{i} c^{\vphantom{\dagger}}_{j}\rangle$, the mean-field Hamiltonian takes the form:
\begin{align}
    H_{\text{MF}}=& -t\sum_{\langle ij\rangle}\left(c^\dagger_{i} c^{\vphantom{\dagger}}_{j}+\mathrm{H.c.}\right)-\mu\sum_{i}c^\dagger_{i} c^{\vphantom{\dagger}}_{i}\nonumber\\
                 & +\frac{1}{2}\sum_{i\neq j}V_{i,j}\Big[\bar{n}_{i}c^\dagger_{j} c^{\vphantom{\dagger}}_{j}+\bar{n}_{j}c^\dagger_{i} c^{\vphantom{\dagger}}_{i}\nonumber\\
                 &\ \ \ \ \ \ \ \ \ \ \ \ \ \ \ \ \  +\Delta_{ji}c^\dagger_{i}c^\dagger_{j}-\psi_{ji}c^\dagger_{i} c^{\vphantom{\dagger}}_{j}+\text{H.c.}\Big]\nonumber\\
                 & -\frac{1}{2}\sum_{i\neq j}V_{i,j}\left[\bar{n}_{i}\bar{n}_{j}-\lvert\psi_{ij}\rvert^2+\lvert\Delta_{ij}\rvert^2\right],
\label{eq_H_MF}
\end{align}
where the pairing field $\Delta_{ij}=-\Delta_{ji}$ is of odd parity to comply with Fermi-Dirac statistics.

The term responsible for phase separation can be seen in Eq.~\eqref{eq_H_MF}.  As discussed in the introduction,  Hartree terms of the form: $\sum_{i\neq j}V_{i,j}\bar{n}_j c^\dagger_{i} c^{\vphantom{\dagger}}_{i}$, effectively renormalize the chemical potential with a density dependence.  The non-linear density dependence gives rise to a negative inverse compressibility, $\kappa^{-1}\sim\partial\mu/\partial n$, which signals phase separation.  At a qualitative level, addition of $V_2$ terms to the $V_1$ terms in Eq.~\eqref{eq_H_V1_V2} serves to "offset" the Hartree renormalization of the chemical potential and therefore keep the inverse compressibility positive.  

To find the possible ordered states of the system, we take random complex numbers as initial values of the mean fields $\bar{n}_{i}$, $\psi_{ij}$, and $\Delta_{ij}$ in Eq.~\eqref{eq_H_MF}.  We then transform the Hamiltonian into momentum space to obtain $\tilde{H}_{\text{MF}}(\mathbf{k})$, where $\mathbf{k}$ is the crystal momentum in the Brillouin zone. By diagonalizing the Hamiltonian in momentum space, we find the eigenenergy and eigenstates at each momentum $\mathbf{k}$ in each band. Then the chemical potential is determined by fixing the fermion number density $n$ through Eq.~\eqref{eq_n}. Using these intermediate results we calculate the  mean fields  $\langle c^\dagger_{i} c^{\vphantom{\dagger}}_{i}\rangle$, $\langle c^\dagger_{i} c^{\vphantom{\dagger}}_{j}\rangle$, and $\langle c^{\vphantom{\dagger}}_{i} c^{\vphantom{\dagger}}_{j}\rangle$, which are used as the new input values in Eq.~\eqref{eq_H_MF} to repeat the above calculation until the values of the mean fields converge. In the self-consistent calculation, we do not impose any particular structure of the mean fields (except for $\Delta_{ij}=-\Delta_{ji}$), but allow them to self-evolve and converge to the lowest-energy state. 

In a honeycomb lattice each site interacts with three NNs and, for $V_2\neq 0$, six NNNs as shown in Fig.~\ref{fig_schematics_honeycomb}(a). To explore possible phases, we choose different unit cells containing $m=2$, $4$, and $6$ sites, respectively, as depicted in Fig.~\ref{fig_schematics_honeycomb}(b). As a result, there are $m$ independent real-valued density fields $\bar{n}_i$, $9m$ independent complex-valued exchange fields $\psi_{ij}$, and $9m$ independent complex-valued pairing fields $\Delta_{ij}$, which must be solved self-consistently. When the system converges to different solutions, for instance resulting from different choices of the unit cell, the solution that minimizes the free energy is accepted.

Our self-consistent mean-field calculation is performed at finite temperatures.  But the temperatures can be taken sufficiently low to extrapolate to the zero temperature limit. For example, in the results shown below, the chemical potential is calculated at low temperatures and then extrapolated to zero temperature.  Here we observe that the temperature dependence of the chemical potential is very weak in the low temperature regime.  Also, here we use the mean field temperature as a proxy for the stability of the topological superfluid with the understanding that a realistic 2D system undergoes a Berezinskii–Kosterlitz–Thouless transition \cite{Berezinsky1972,Kosterlitz1973} with increasing temperature.

\section{Competing Orders}
\label{sec_phases}

%%%%%%%%%%%%%%%%%%%%%%%%%%%%%%%%%%%%%%%%%%%%%%%
\begin{figure}[ht]
	\begin{center}
	   %\vspace{1cm}
		\includegraphics[width=0.48\textwidth]{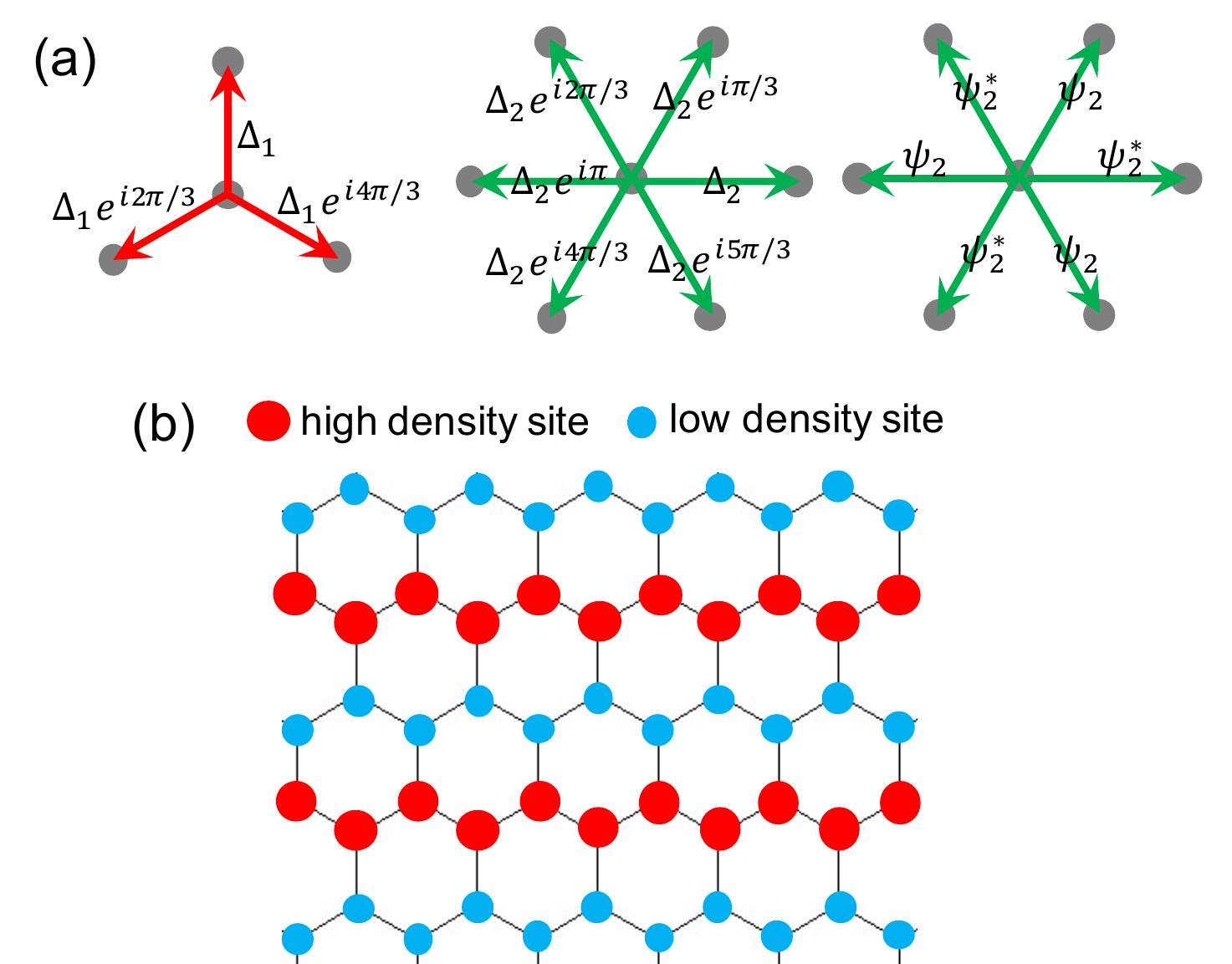}
	\end{center}
	\caption{ (a) Bond order parameter pattern in the superfluid phase.  The left red (middle green) arrows depict the pairing order parameter along NN (NNN) bonds in the honeycomb lattice. And the right green arrows show the exchange field along the NNN bonds when it has a finite imaginary part. (b) The zigzag stripe phase pattern in the honeycomb lattice. }
	\label{fig_schematic_orders}
\end{figure}
%%%%%%%%%%%%%%%%%%%%%%%%%%%%%%%%%%%%%%%%%%%%%%%

%%%%%%%%%%%%%%%%%%%%%%%%%%%%%%%%%%%%%%%%%%%%%%%
\begin{figure}[ht]
	\begin{center}
	   %\vspace{1cm}
		\includegraphics[width=0.48\textwidth]{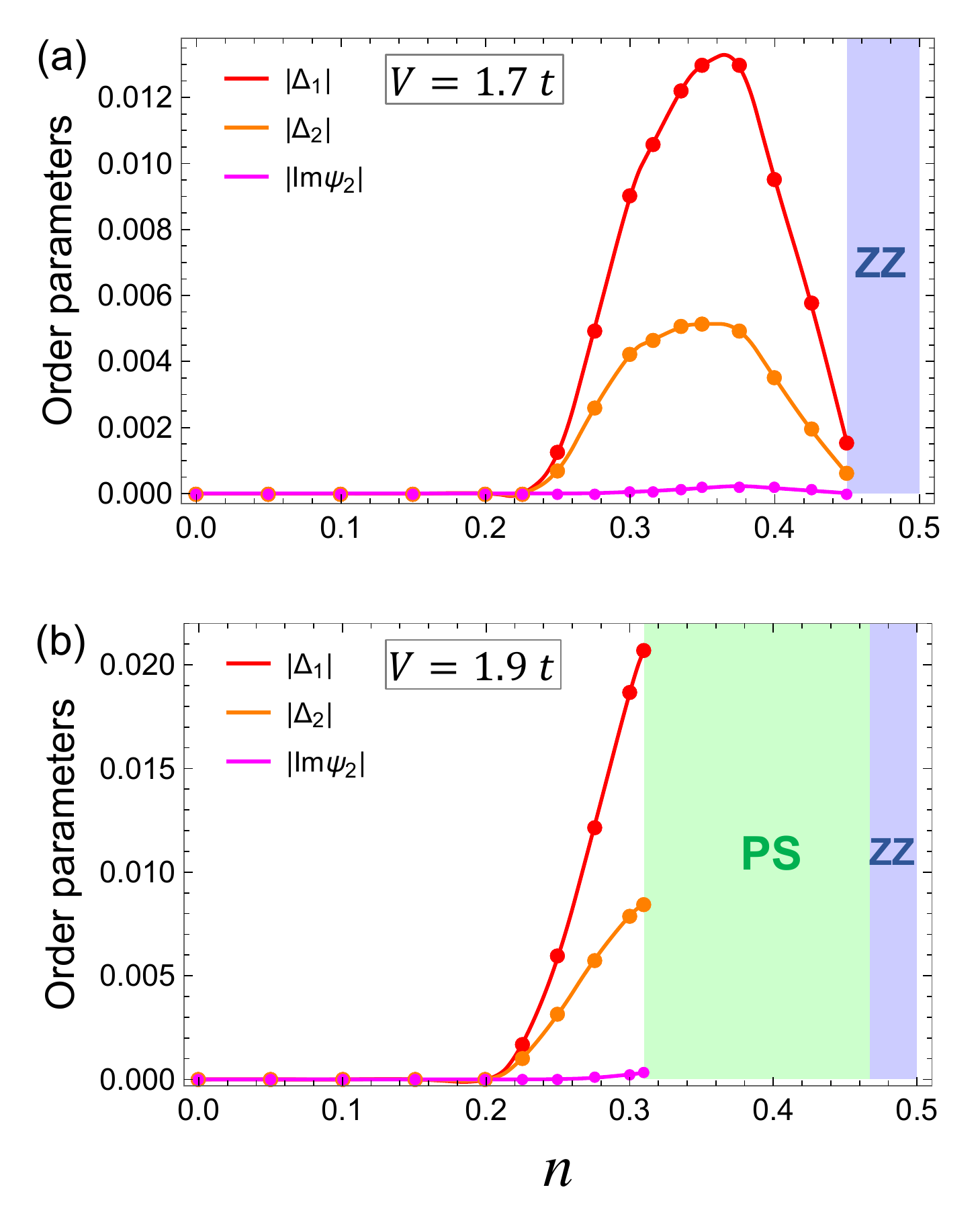}
	\end{center}
	\caption{The red circles plot the low temperature ($T=0.001t$) magnitudes of the order parameters in the superfluid phase in the cases of $V_1=-V$ and $V_2=V/3$ with (a) $V=1.7t$ and (b) $V=1.9t$ as a function of density. The lines are guides to the eye that refer to the magnitudes of the NN pairing fields $|\Delta_1|$, the NNN pairing fields $|\Delta_2|$, and the imaginary part of the NNN exchange fields $|\mathrm{Im}\psi_2|$ at filling fractions $n\leq 0.5$. The results at $n>0.5$ are the same as those at $(1-n)$. The Van Hove singularity is at filling 3/8. }
	\label{fig_calculated_orders}
\end{figure}
%%%%%%%%%%%%%%%%%%%%%%%%%%%%%%%%%%%%%%%%%%%%%%%

The Hartree-Fock equations reveal several competing orders but care must be taken in analyzing the underlying phase diagram because of phase separation.  We find the following phases as we take different limits of our model: a normal metal phase ($\Delta_{ij}=\mathrm{Im}\psi_{ij}=0$), a superfluid ($\vert \Delta_{ij}\vert>0$ on NN and/or NNN bonds) sometimes with very weak chiral currents ($\mathrm{Im}\psi_2\neq 0$, where $\psi_2$ is the exchange field $\psi_{ij}$ on NNN bonds), a zigzag stripe phase (where $\bar{n}_i$ oscillates spatially), and phase separation.  We note that a non-zero  $\mathrm{Re}\psi_{ij}$ renormalizes the dispersion relation but does not lead to phase transition in this work.  In this section we discuss these orders in detail. 

We first discuss the uniform superfluid case that can arise from just NN attraction, i.e., $V_1=-V$ and $V_2=0$ for $V>0$ in Eq.~(\ref{eq_H_V1_V2}).   In this work we consider weak to intermediate interaction strengths $V\lesssim 2.0t$.  For weak interactions, at all particle densities but half-filling, $0<n<1$ and $n\neq 0.5$, the system develops a pairing instability towards the uniform superfluid phase with chiral $p$-wave symmetry at low temperatures.  The red arrows in Fig.~\ref{fig_schematic_orders}(a) depict the complex pairing order parameter on NN bonds.  
The pairing order parameter we find has a phase structure with relative phase on different bonds determined by the relative bond orientations, i.e., the order parameter on the NN bonds has the form of $\Delta_1 e^{\pm il2\pi/3}$ where $l=0,1,2$ and $\Delta_1$ is a complex number with an arbitrary global phase.  Therefore, the pairing order is of chiral $p$-wave symmetry and the chirality $\pm 1$ is spontaneously chosen.  As a result of the pairing instability, the spectrum is fully gapped on the Fermi surface in the superfluid phase.  (In the half-filling $n=0.5$ case the system does not open a gap and remains a semimetal at all temperatures.)  

The addition of repulsive NNN interactions ($V_2>0$) leads to the possibility of density-wave order that competes with the chiral superfluid.  The dominant effect of NNN repulsion is to drive the formation of a zigzag stripe phase near half filling. Unlike the uniform superfluid phase where there is no density order, i.e., $\bar{n}_i=n$, the zigzag stripe phase has alternating particle number occupation of the high-density sites $n_h=n+\delta n$ denoted by the bigger red dots in Fig.~\ref{fig_schematic_orders}(b), and the low-density sites $n_l=n-\delta n$ represented by the smaller blue dots. $\delta n$ reaches a maximum at half filling. The same-density sites form stripes along one of the six zigzag directions which lead to a six-fold degeneracy of the zigzag stripe state. In the zigzag phase, at very low temperatures we find a residual anisotropic (non chiral) $p$-wave pairing order with maximal pairing amplitude along the high-density stripes and zero pairing amplitude perpendicular to the stripes, but this $p$-wave pairing vanishes quickly with increasing temperature.  In the following we ignore the weak pairing order in the stripe phase.  The critical temperature of the zigzag stripe phase itself is much higher than that of the uniform topological superfluid phase. 

The superfluid phase established in the system with NN attractions and NNN repulsion also exhibits NNN pairing.  In such a case the superfluid phase is characterized not just by finite pairing fields on the NN bonds, but also on NNN bonds as shown by the green arrows in the middle of  Fig.~\ref{fig_schematic_orders}(a).  Here, the pairing order parameter on the NNN bonds is $\Delta_2 e^{\pm il'\pi/3}$ with $l'=0,1,..,5$ and $\Delta_2$ a complex number ($\Delta_2$ also has a relative phase with respect to $\Delta_1$ determined by the NN and NNN bond orientation difference). 

Figures~\ref{fig_calculated_orders}(a)-(b) show the calculated pairing magnitudes $|\Delta_1|$ and $|\Delta_2|$ at $T=0.001t$ for $V_1=-V$ and $V_2=V/3$ in the cases of $V=1.7t$ and $V=1.9t$. The results at $n>0.5$ are the same as those at $(1-n)$ due to particle-hole symmetry. In addition to the pairing fields, there is also a small but non-zero imaginary part of the exchange field on the NNN bonds $\mathrm{Im}\psi_2\neq 0$, indicating a small bond current generated by the nontrivial Berry curvature of the band. The pattern of the NNN exchange fields is shown on the green arrows on the right of Fig.~\ref{fig_schematic_orders}(a). Note that both the pairing field and the exchange field depend on the bond direction: $\Delta_{ij}=-\Delta_{ji}$ and $\psi_{ij}=\psi_{ji}^*$. Both the chiral $p$-wave pairing and the tiny loop current order exist in the topological superfluid phase as allowed by time-reversal symmetry breaking.

We now discuss phase separation.  We start with the purely attractive case, i.e., $V_2=0$.  At fixed particle density $n\,(n\neq 0.5)$, increasing the attraction strength $V_1$ leads to higher critical temperature $T_c$ of the superfluid phase. And at fixed attraction strength, the pairing order is enhanced by larger filling fractions $n$ and reaches its maximum around the Van Hove filling. For $0<n<0.5$, one would expect that the critical temperature rises with increasing $n$ and maximizes around the Van Hove fillings before decreasing to zero at half filling. (The case at $0.5<n<1$ is the same as that at $1-n$.) However, when interactions are large enough, the system becomes dynamically unstable starting in the neighborhood of the Van Hove fillings and expanding to the whole filling regime when above a critical interaction strength. This instability is characterized by the non-monotonic dependence of the chemical potential $\mu$ on the particle density $n$. In the case with NNN repulsion added ($V_2>0$), the competition from the density-wave order can also lead to a non-monotonic $\mu$-$n$ dependence at large interactions.  We must therefore also consider the possibility of phase separation. 

To define the phase separated region in parameter space, we must compute the free energy gain in forming the phase separated state.  We quantify phase separation into a mixture of two domains with different densities $n_1$ and $n_2$ using Maxwell construction. Note that the two domains can have either the same order (as in the case of NN attractions only) or different orders (as in the case with NNN repulsion added). The density region $(n_1,n_2)$ of phase separation is determined as follows \cite{Kapcia2016}.  For a system with fixed particle density $n$ ($n_1<n<n_2$), in the state of phase separation the system separates into two domains with particle density $n_1$ and $n_2$, respectively,
\begin{equation}
    n=\alpha n_1+(1-\alpha)n_2,
    \label{eq_PS_density}
\end{equation}
where $\alpha$ is the fraction of the domain with particle density $n_1$ while $(1-\alpha)$ is the fraction of the domain with $n_2$. And the free energy density (the free energy per site) of the phase separated state is:
\begin{equation}
    f_{\text{PS}}(n_1,n_2)=\alpha f_1(n_1)+(1-\alpha)f_2(n_2),
    \label{eq_PS_free_energy}
\end{equation}
where $f_i(n_i)$ ($i=1,2$) is the free energy density of a uniform phase with particle density $n_i$. The value of $f_{\text{PS}}$ is determined by minimizing (\ref{eq_PS_free_energy}) with respect to $n_1$ and $n_2$. This, together with (\ref{eq_PS_density}), yields the equilibrium condition:
\begin{align}
    \mu_1(n_1)&=\mu_2(n_2)=\mu_{\text{PS}},\label{eq_mu1_mu2}\\
    f_1(n_1)-\mu_1 n_1&=f_2(n_2)-\mu_2 n_2,
    \label{eq_f1_f2}
\end{align}
where $\mu_{\text{PS}}$ is the chemical potential of the phase separated state which is independent of the particle density $n$, i.e., $\partial\mu_{\text{PS}}/\partial n=0$. Eqs.~\eqref{eq_mu1_mu2}-\eqref{eq_f1_f2} are the conditions used to identify the values of $n_1$ and $n_2$ for the densities with phase separation.

\section{Enhancing Topological Superfluidity}
\label{sec_enhancing}

\begin{figure}[ht]
	\begin{center}
	   %\vspace{1cm}
		\includegraphics[width=0.48\textwidth]{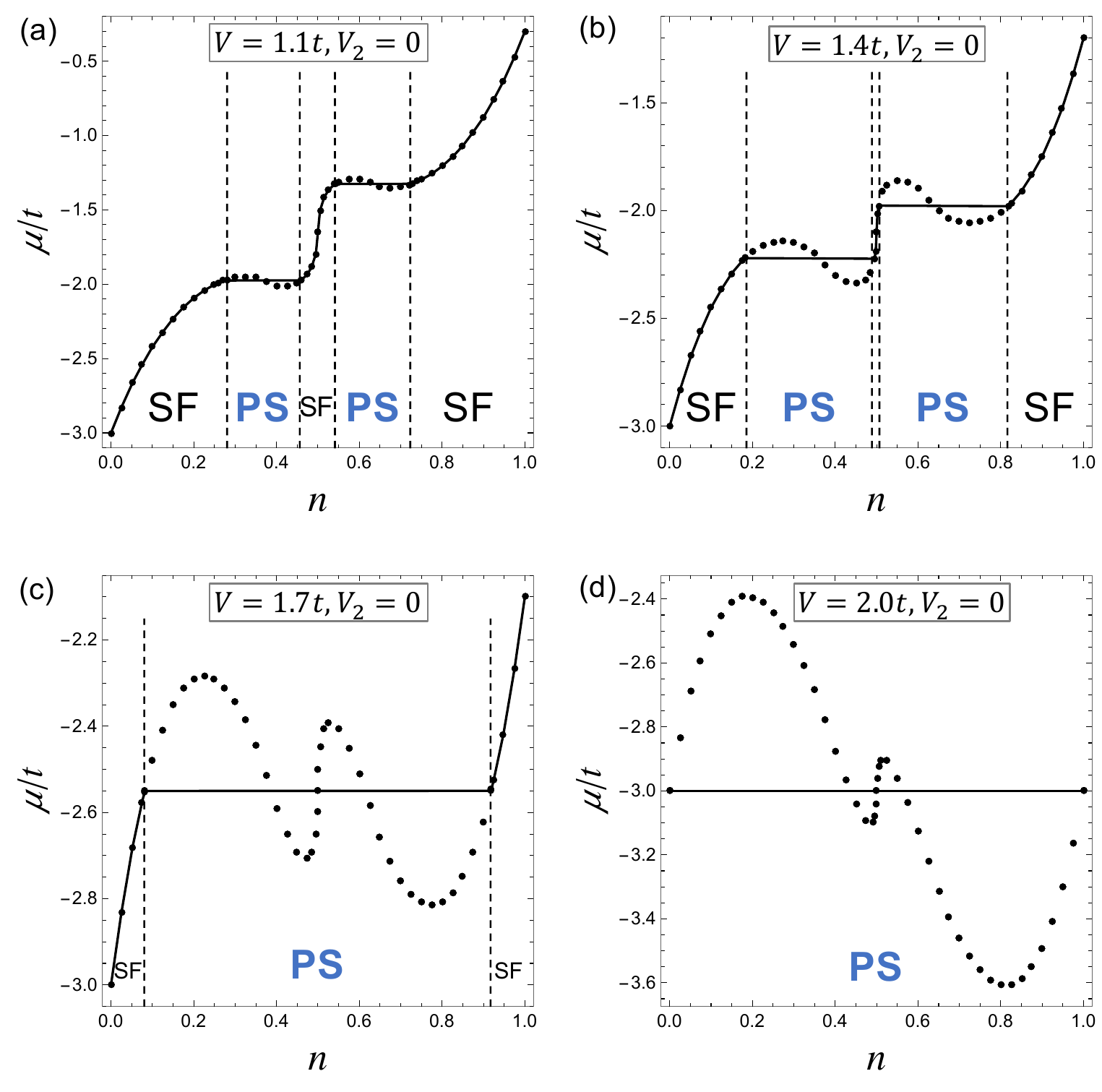}
	\end{center}
	\caption{The circles plot results from our calculation of the chemical potential $\mu$ against particle number density $n$ extrapolated to zero temperature. The lines are a guide to the eye.  A flat horizontal line indicates a phase separated region.  The interaction is chosen to be attractive between nearest neighbors, $V_1=-V$ and $V_2=0$, where (a) $V=1.1t$, (b) $V=1.4t$, (c) $V=1.7t$, and (d) $V=2.0t$.  We see that increasing the strength of the attractive interaction leads to phase separation (PS) over a larger range of densities and superfluid (SF) over a smaller range of densities.   }
	\label{fig_mu_n_zeroV2}
\end{figure}

We can now discuss how to systematically strengthen topological superfluidity.  As discussed above, increased attraction $V_1$ leads to phase separation for $V_2=0$.  In the following we show that $V_2>0$ helps stabilizing superfluidity to a certain degree to allow higher temperature superfluids.

We first demonstrate that only NN attraction [$V_1=-V$ and $V_2=0$ for $V>0$ in Eq.~\eqref{eq_H_V1_V2}] does not have a strong superfluid phase. 
Fig.~\ref{fig_mu_n_zeroV2} shows the calculated values of the chemical potential (the dotted lines) at varying densities at $T=0$ (extrapolated from low temperature results) with the interaction strength $V=1.1t$ in Fig.~\ref{fig_mu_n_zeroV2}(a), $V=1.4t$ in Fig.~\ref{fig_mu_n_zeroV2}(b), $V=1.7t$ in Fig.~\ref{fig_mu_n_zeroV2}(c), and $V=2.0t$ in Fig.~\ref{fig_mu_n_zeroV2}(d).  The chemical potential varies continuously with the particle density and the inverse compressibility $\kappa^{-1}\sim\partial\mu/\partial n$ is well defined. The appearance of the negative inverse compressibility, i.e., the chemical potential becomes a decreasing function of the density, in the neighborhood around the Van Hove filling in Fig.~\ref{fig_mu_n_zeroV2}(a) signals that the system is unstable towards phase separation.  The density region $n_1<n<n_2$ of phase separation can be deduced by the Maxwell construction as discussed above. 

%%%%%%%%%%%%%%%%%%%%%%%%%%%%%%%%%%%%%%%%%%%%%%%
\begin{figure}[ht]
	\begin{center}
	   %\vspace{1cm}
		\includegraphics[width=0.48\textwidth]{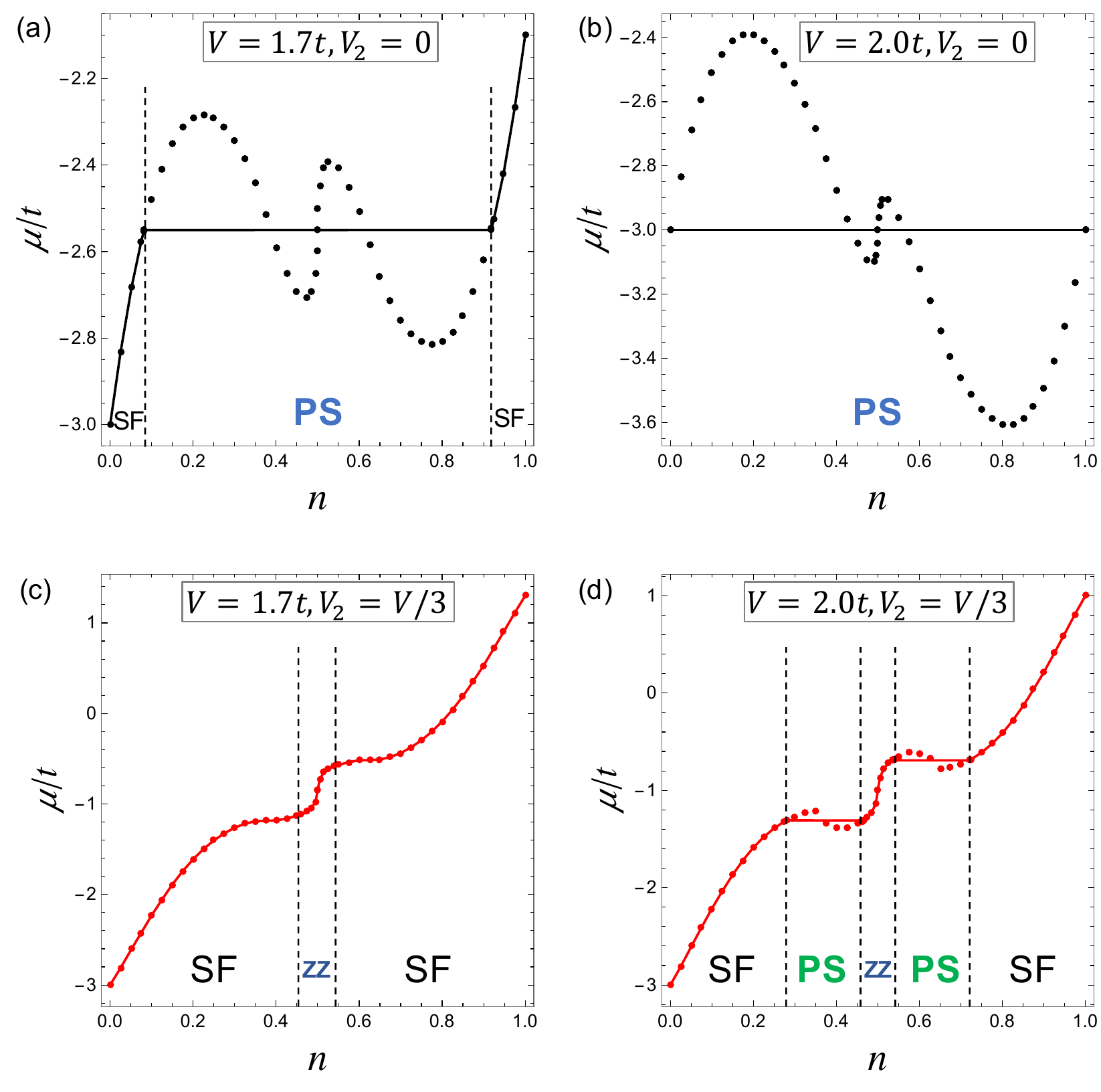}
	\end{center}
	\caption{The same as Fig.~\ref{fig_mu_n_zeroV2} where (a) and (b) are cases with $V_1=-V$ and $V_2=0$. (c) and (d) are cases with $V_1=-V$ and $V_2=V/3$. By comparing the top row to the bottom row we see that the next nearest neighbor repulsion suppresses PS in favor of SF.  A stripe zigzag phase (ZZ) also appears near $n=0.5$. }
	\label{fig_mu_n_thirdV2}
\end{figure}
%%%%%%%%%%%%%%%%%%%%%%%%%%%%%%%%%%%%%%%%%%%%%%%

Here the system separates into two domains of different particle densities $n_1$ and $n_2$, each associated with a lower superfluid critical temperature, i.e., $T_c(n_1)<T_c(n)$ and $T_c(n_2)<T_c(n)$. And the chemical potential of the mixture remains constant from $n_1$ to $n_2$, as indicated by the flat line segments in the $\mu$-$n$ plots. With increasing interaction strengths as shown in Fig.~\ref{fig_mu_n_zeroV2}(b)-(d), the phase separation region is enlarged and eventually the system becomes completely phase separated at all densities displayed in Fig.~\ref{fig_mu_n_zeroV2}(d). It is this instability towards phase separation that limits the ability to increase the critical temperature by increasing interaction strengths or increasing the density of states. In the honeycomb system with just NN attraction, we find that the highest critical temperature $T_c^{\text{NN}}$ of a uniform topological superfluid phase is very low, on the order of $0.001t$.

We now discuss the addition of NNN repulsion to stabilize topological superfluidity, i.e., $V_2>0$.  Fig.~\ref{fig_mu_n_thirdV2} shows the comparison of the calculated chemical potential at varying densities at $T=0$ in the case of $V_1=-V, V_2=0$ [black dotted lines in Fig.~\ref{fig_mu_n_thirdV2}(a)-(b)] and in the case of $V_1=-V, V_2=V/3$ [red dotted lines in Fig.~\ref{fig_mu_n_thirdV2}(c)-(d)] for $V=1.7t$ and $V=2.0t$, where the superfluid phase in the latter case has chiral $p$-wave symmetry as well.

Although the presence of longer-range repulsive interaction weakens the pairing strength at a certain density, it more efficiently reduces the phase separation and sustains the uniform superfluid phase to higher densities at larger interactions. 
Comparison between Figs.~\ref{fig_mu_n_thirdV2}(a)-(b) and  Figs.~\ref{fig_mu_n_thirdV2}(c)-(d) shows that the density region of a uniform topological superfluid phase is significantly enlarged by the addition of the weaker NNN repulsion $V_2$.  

Another effect from the additional NNN repulsion is the emergence of zigzag stripe density-wave order around half filling when the interaction is large enough. This density wave order competes with the topological superfluidity. As shown in Fig.~\ref{fig_mu_n_thirdV2}(c), there is a direct first order phase transition from the chiral $p$-wave pairing order to the zigzag stripe order when increasing the particle density towards half filling at $V=1.7t$ or below. 

Further increasing the interaction strength $V$ results in a non-smooth change in the chemical potential.  The chemical potential drops right above a critical density $n_c$, associated with a transition from the chiral $p$-wave pairing state to the zigzag stripe state, and continues to decrease slightly before increasing again. At large interactions, the drop in chemical potential becomes substantially discontinuous as shown in Fig.~\ref{fig_mu_n_thirdV2}(d) by the red dots, where $V=2.0t$ and the calculated $\mu$ value drops discontinuously right above $n_c=0.35$ and $0.65$.  Meanwhile the system transits from the chiral $p$-wave order to the zigzag stripe. 

We therefore see that, at $V=2.0t$ as shown in Fig.~\ref{fig_mu_n_thirdV2}(d), the chiral $p$-wave pairing phase is dominant at lower densities, but with increasing density there appears a density region with phase separation where the chemical potential $\mu_{\text{PS}}$ remains constant (the red flat line segment) and the system becomes a mixture of topological superfluid and zigzag stripe domains. The system moves into the zigzag stripe phase when further increasing the density towards half filling.  The zigzag stripe order is the strongest at half filling $n=0.5$ \cite{Capponi2015}. Although in the case of $V=2.0t$ the density region of the uniform topological superfluid phase is smaller than that in the case of $V=1.7t$, the pairing strength at the same density is stronger in the former due to its larger interaction strength. Hence, the search for the optimal values of interactions needs to take into account both pairing strength and pairing density.

As illustrated above, the addition of a weaker NNN repulsion to the NN attraction suppresses the phase separation instability in favor of the topological superfluid phase to a certain degree.  To further optimize parameters, we  perform a systematic study for a fixed value of $V=2.0t$ ($V_1=-V$) by varying the ratio $V_2/V$ (we choose $V_2/V=1/4$, 1/3, 2/5, 1/2, and 3/5).  We find that $V_2$ can not be too weak in order to prevent the phase separation instability resulting from the Hartree effect of the attractive interaction $V_1$. But $V_2$ can not be strong because its repulsive nature weakens the pairing strength and its induced density wave order (zigzag stripe) competes with superfluidity. The optimal value we find is $V_2/V \approx 1/3$.

We also search for optimal values of $V$. We set  a fixed ratio $V_2/V=1/3$ while varying $V/t$ (We choose $V/t=1.7$, 1.8, 1.9, 2.0, and 2.1).  We find that weaker $V$ lowers the pairing strength. But a too strong $V$ gives rise to a large phase separation region and reduces the pairing density significantly. Taking into account both the pairing strength and pairing density, we find that the optimal choice of $V$ and $V_2/V$ values is around $V/t=1.9$ and $V_2/V=1/3$.

%%%%%%%%%%%%%%%%%%%%%%%%%%%%%%%%%%%%%%%%%%%%%%%
\begin{figure}[ht]
	\begin{center}
	   %\vspace{1cm}
		\includegraphics[width=0.48\textwidth]{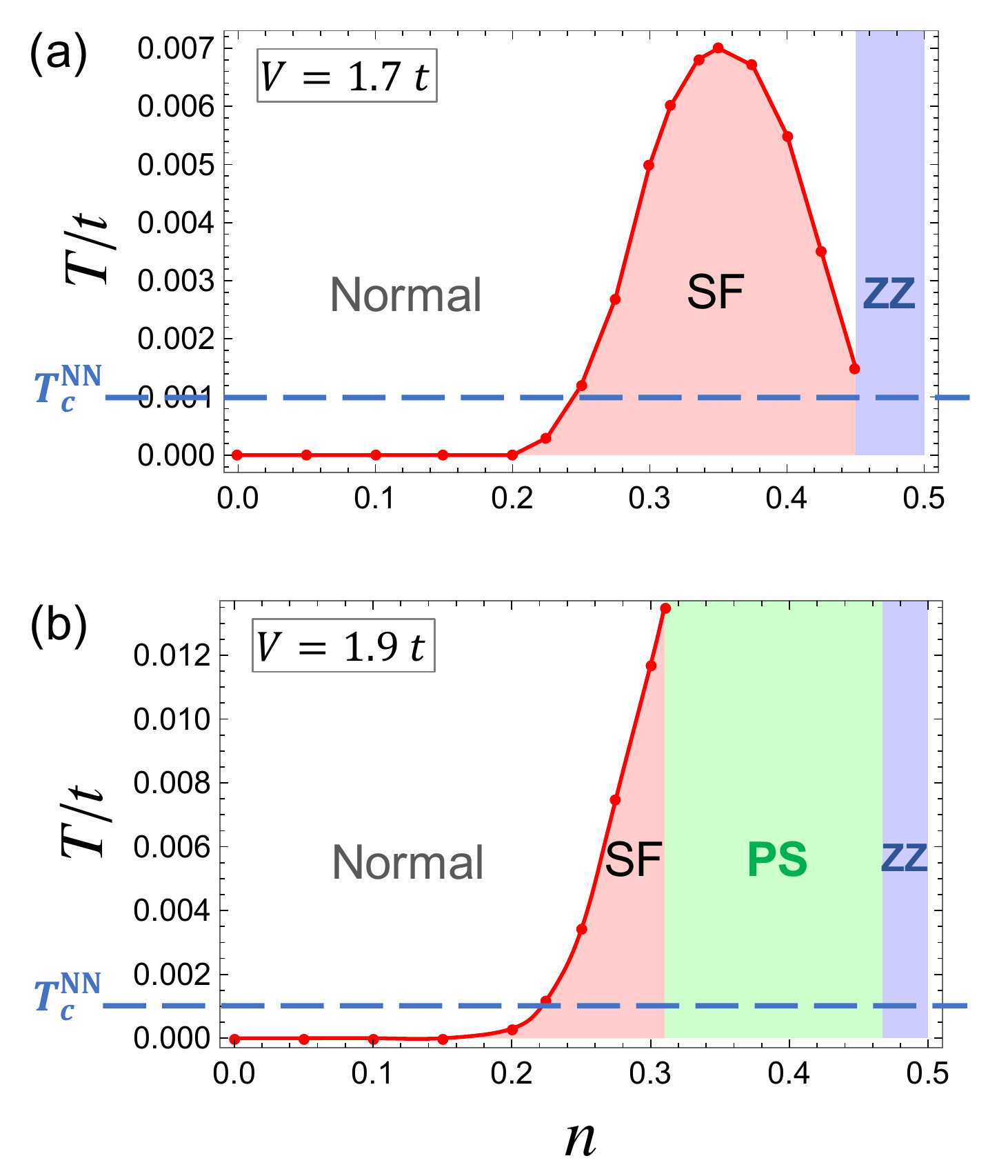}
	\end{center}
	\caption{Calculated $T$-$n$ Phase Diagrams. The red circles refer to the critical temperature $T_c$ of the chiral $p$-wave pairing phase obtained from our calculation.  The solid red lines are a guide to the eye. The results at $n>0.5$ are the same as those at $(1-n)$. Here we have chosen interaction parameters $V_1=-V$ and $V_2=V/3$ with (a) $V=1.7t$ and (b) $V=1.9t$ to optimize the critical temperature of the superfluid.  We find that the case with $V=1.9t$ leads to a significant increase in critical temperature near $n\approx 0.3$.  The blue dashed line shows, for comparison, the highest critical temperature found for just NN attraction and no NNN repulsion.       }
	\label{fig_T_n_phase_diagram}
\end{figure}
%%%%%%%%%%%%%%%%%%%%%%%%%%%%%%%%%%%%%%%%%%%%%%%

Figure~\ref{fig_T_n_phase_diagram} shows the  calculated $T$-$n$ phase diagrams at $n\leq 0.5$ with $V_1=-V$ and $V_2=V/3$, where the red connected dotted lines refer to the critical temperature $T_c$ of the topological superfluid phase. The results at $n>0.5$ are the same as those at $(1-n)$. In the case of $V=1.7t$ as shown in Fig.~\ref{fig_T_n_phase_diagram}(a) at low temperatures with increasing particle density $n$ the system first experiences a second-order phase transition from the normal (metallic) phase to the superfluid phase, and then a direct first-order phase transition from the uniform topological superfluid phase, denoted by the red shaded area, to the zigzag stripe phase, denoted by the blue shaded area, when the density approaches half filling $n=0.5$. And the highest critical temperature is approximately $T_c\approx 0.007 t$. In the case of $V=1.9t$ as shown in Fig.~\ref{fig_T_n_phase_diagram}(b), instead of a direct first-order phase transition from the superfluid to the zigzag stripe phase, a phase separated region denoted by the green shaded area occurs indicating the system becomes a mixture of two phases before moving into the zigzag stripe phase near half filling.  Larger interaction strengthens the pairing such that the highest critical temperature in this case is around $T_c\approx 0.013 t$. In both cases, the maximal critical temperatures are much higher than that in the case of $V_2=0$, i.e., $T_c^{\text{NN}}\approx 0.001 t$. 

We have also performed calculations including the third- and fourth-neighbor interactions to study the impact of a long-range tail. For weak enough repulsive long-range tail the topological superfluid phase is not affected, and is even more robust if the long-range tail is of the RKKY type, i.e., having alternating interaction sign change with distance.  

\section{Discussion and Conclusion}
\label{sec_conclusion}

Our studies identify a mechanism to enhance the stability of a topological superfluid made from spinless lattice fermions.  We argue that the addition of weaker longer-range repulsion to the short-range attraction helps stabilize the topological superfluid phase against phase separation at high density of states and large interaction strength. Our unrestricted Hartree-Fock  calculation for a honeycomb lattice reveals an order of magnitude enhancement of the topological superfluidity critical temperature, making this intriguing phase more experimentally accessible. For the honeycomb lattice, the addition of the longer-range repulsive interactions also introduces a competing zigzag stripe order which becomes dominant near half filling at large interactions, consistent with exact diagonalization studies \cite{Capponi2015,Capponi2017}. Although the competition from the zigzag phase at larger interactions leads to the appearance of another type of phase separation at certain densities, the superfluid order can still benefit from the increased interaction strength thus allowing a higher critical temperature.

We expect that our results apply to other lattices.  This expectation is based on the observation that the Hartree effect driving phase separation is a generic mean-field shift effectively adding density dependence to a renormalized chemical potential.  On a qualitative level, the addition of longer-range repulsion tends to roughly cancel the Hartree effect thus allowing the topological superfluid to persist.  Future work will quantitatively explore this mechanism in other lattices where we expect \cite{Midtgaard2016} even higher critical temperatures.

%%-------------------------------------------------------
\section*{Acknowledgments}

We acknowledge support from AFOSR (FA9550-18-1-0505,FA9550-19-1-0272), ARO (W911NF2010013), and NSF 2014157.

\bibliography{references}

\end{document}